\documentclass{PoS}

\usepackage{epsfig}

\def\slash#1{\setbox0=\hbox{$#1$}\dimen0=\wd0
      \setbox1=\hbox{/} \dimen1=\wd1 \ifdim\dimen0>\dimen1
      \rlap{\hbox to \dimen0{\hfil/\hfil}} #1                        \else
      \rlap{\hbox to \dimen1{\hfil$#1$\hfil}}
      /   \fi}
\def\simge{\mathrel{\rlap{\raise 0.511ex \hbox{$>$}}{\lower 0.511ex
\hbox{$\sim$}}}}
\def\simle{\mathrel{\rlap{\raise 0.511ex \hbox{$<$}}{\lower 0.511ex
\hbox{$\sim$}}}}
\def\slash#1{\setbox0=\hbox{$#1$}\dimen0=\wd0
      \setbox1=\hbox{/} \dimen1=\wd1 \ifdim\dimen0>\dimen1
      \rlap{\hbox to \dimen0{\hfil/\hfil}} #1                        \else
      \rlap{\hbox to \dimen1{\hfil$#1$\hfil}}
      /   \fi}

\newcommand{\be}{\begin{equation}}
\newcommand{\ee}{\end{equation}}
\newcommand{\bea}{\begin{eqnarray}}
\newcommand{\eea}{\end{eqnarray}}

\newcommand{\beq}{ \begin{equation}}
\newcommand{\eeq}{ \end{equation}}
\newcommand{\beqau}{\begin{eqnarray*}}
\newcommand{\eeqau}{\end{eqnarray*}}
\newcommand{\beqa}{\begin{eqnarray}}
\newcommand{\eeqa}{\end{eqnarray}}

\newcommand{\Lam}{\Lambda_{\rm QCD}}

\newcommand{\ms}{{\overline{\rm{MS}}}}
\newcommand{\as}{\alpha_{ s}}

\newcommand{\gev}{\,{\rm GeV}}

\newcommand{\condval}{-(265\pm 5_{stat}\pm 22_{syst}\,\textrm{MeV})^3}
\PoS{PoS(LAT2005)081}

\title{Operator product expansion and quark condensate from LQCD in coordinate space}

\ShortTitle{Operator product expansion and quark condensate from LQCD in coordinate space }

\author{Vincente Gim\'enez\\
        Universidad Val\`encia and IFIC\\
        E-mail: \email{vincente.gimenez@uv.es}}

\author{Vittorio Lubicz\\
        Universit\`{a} di Roma Tre and INFN\\
        E-mail: \email{lubicz@fis.uniroma3.it}}

\author{Federico Mescia\\
        Universit\`{a} di Roma Tre and LFN\\
        E-mail: \email{mescia@fis.uniroma3.it}}

\author{\speaker{Valentina Porretti}\\
        Universit\`{a} di Roma Tre and INFN\\
        E-mail: \email{porretti@fis.uniroma3.it}}

\author{Juan Reyes\\
        Universit\`{a} di Roma La Sapiennza\\
        E-mail: \email{juan.reyes@roma1.infn.it}}

\abstract{We perform an exploratory study of the operator product expansion of the quark propagator on the
lattice at short distance in coordinate space. This permits a simple determination of the quark condensate,
$\langle\bar \psi \psi \rangle^{\ms}(2 \textrm{GeV})=-(265\pm 5_{stat}\pm 22_{syst}\,\textrm{MeV})^3 $, and
of the renormalization constant of the quark field, $Z_\psi^{\ms}(2\gev) = 0.871\pm 0.003_{stat}\pm 0.020_{syst}$. This new
method also provides a remarkable non-perturbative test of the OPE predictions at short distance in QCD.}

\FullConference{XXIIIrd International Symposium on Lattice Field Theory\\

                 25-30 July 2005\\

                 Trinity College, Dublin, Ireland}

\begin{document}

\section{Introduction}
 The chiral quark condensate (QC) $ \langle\bar\psi \psi \rangle $
 plays a central role in the non-perturbative sector of QCD. It is expected to be the order parameter which
 controls
the spontaneous chiral symmetry breaking and sets the scale of the pion masses, provided its value is
non-vanishing and its order of magnitude appropriate. To verify these basic requirements, high precision
estimates of the QC, though welcome, are not necessary.

 At the same time there is no easy
experimental access to the QC, as some theoretical assumptions always enter. This suggests to look for
different methods based on independent hypothesis which, to some extent and indirectly, can be tested in this
way. Two examples are the classical measures of the QC extracted from the pion scattering length through chiral
perturbation theory ~\cite{Colangelo:2001sp} or from the nucleon and $B^*-B$ mass-splitting sum rules ~\cite{Dosch:1997wb}.

On the other hand lattice QCD can give an estimate of the QC based on first principles. The simplest approach to compute the QC
on the lattice uses the GMOR formula, which is based on the axial chiral Ward identity
~\cite{Giusti:1998wy}. Alternative determinations of the quark condensate on the lattice have been 
obtained in the framework of the $\epsilon$-expansion of QCD in a small volume ~\cite{Hernandez:1999cu} and from the study
 of the Goldstone
pole contribution to the pseudoscalar quark Green function ~\cite{Becirevic:2004qv}.

The method proposed in ~\cite{Gimenez:0503001} and summarized in this talk starts from the operator product expansion
of the quark propagator at short euclidean distance: \be S(x^2) \;= \;\frac{1}{2\pi^2}C_I(x^2) \,
\frac{\slash x}{(x^2)^2}\; +\; \frac{1}{4\pi^2}C_m(x^2)\, \frac{m}{x^2} \; -\frac{1}{4 N_C} \; C_{\bar \psi
\psi}(x^2)\, \langle\bar\psi\psi\rangle\; +\; \ldots \label{eq:ope} \ee
 The quark propagator on the l.h.s. is computed on the lattice and the QC is
extracted fitting the r.h.s. in the chiral limit. Our final result reads $\langle\bar \psi
\psi\rangle^{\ms}(2\, \textrm{GeV})=-(265\pm 23\,\textrm{MeV})^3 $.

 The idea is simple and provides a non-perturbative
test of the hypothesis behind, namely the validity of the operator product expansion in QCD. The precision
achieved in this exploratory study is not high, but  can be improved in time, being mainly computer-limited.
However the method itself is flexible and powerful: in priciple, from the fit of eq.~(\ref{eq:ope}), the
quark masses and the gluon condensate which appears at ${\cal O}(x^2)$ can be obtained too. Another advantage
is that the renormalization procedure is rather simple: once the quark propagator is renormalized, its
operator product expansion is expressed in terms of renormalized quantities too. The same holds for the
$O(a)$ improvement.

\section{Lattice simulation}\label{2}
We generated 180 gauge configurations in the quenched approximation with the $O(a)$-improved Wison action, on
a volume $32^3$x70 at $\beta=6.45$. This value of the coupling corresponds to an inverse lattice spacing
$a^{-1}=3.87(19) \gev$, evaluated from the $K$ and $K^*$ meson masses with the lattice plane method. The
simulated values of the hopping parameter are $\kappa=$ 0.1349, 0.1351, 0.1352, 0.1353 and correspond to
light quark masses in the range $m_s/2 \simle m \simle m_s$. The values of the renormalized masses used in
the chiral extrapolations and in eq.~(\ref{eq:ope}) have been taken from ~\cite{masse}. The statistical
errors are computed with the jackknife technique.
\begin{figure}[t!]
\begin{tabular}{cc}\vspace{-1.0cm}
\epsfxsize5.7cm\epsffile{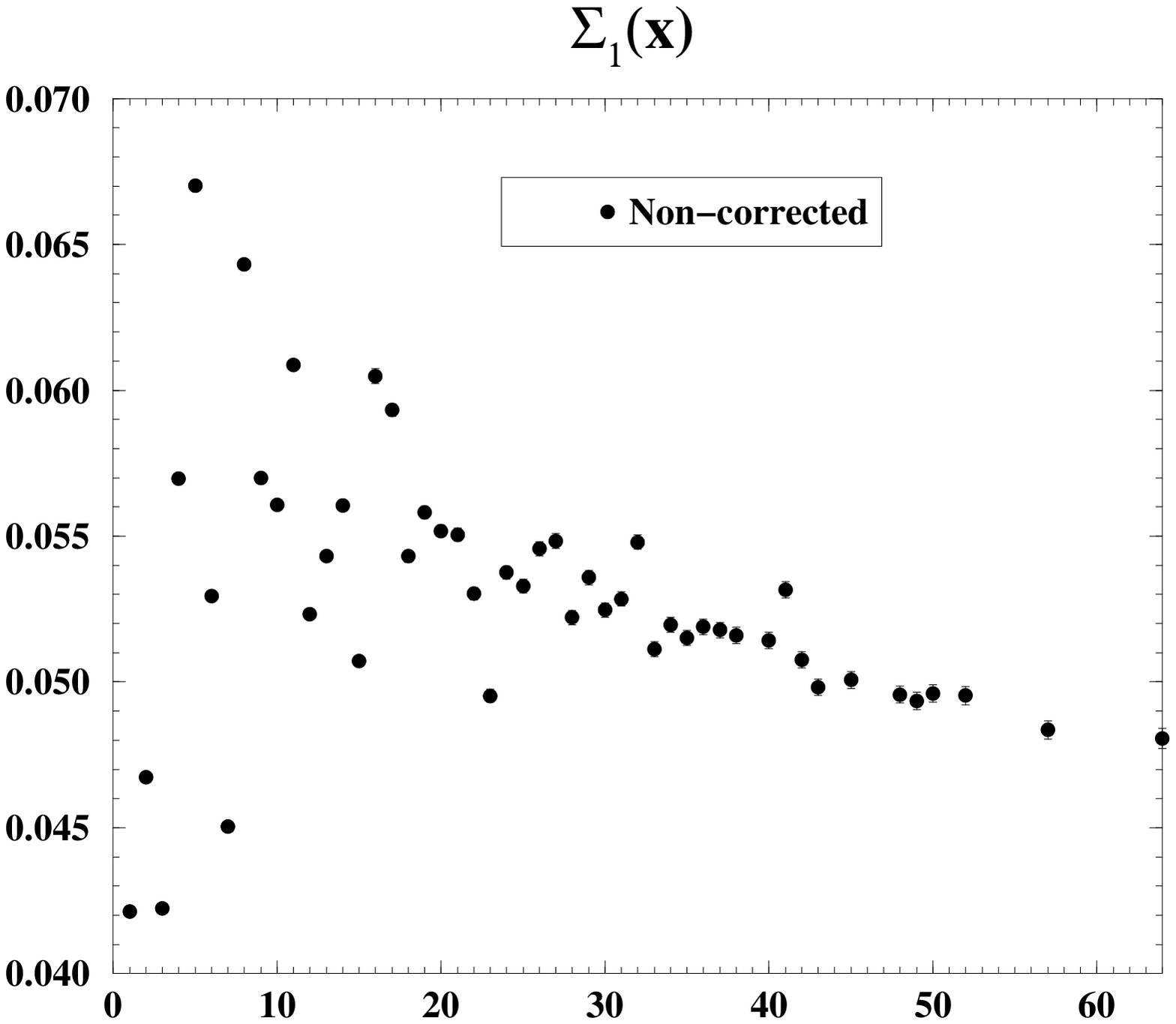} & \hspace*{0.5cm} \epsfxsize5.7cm\epsffile{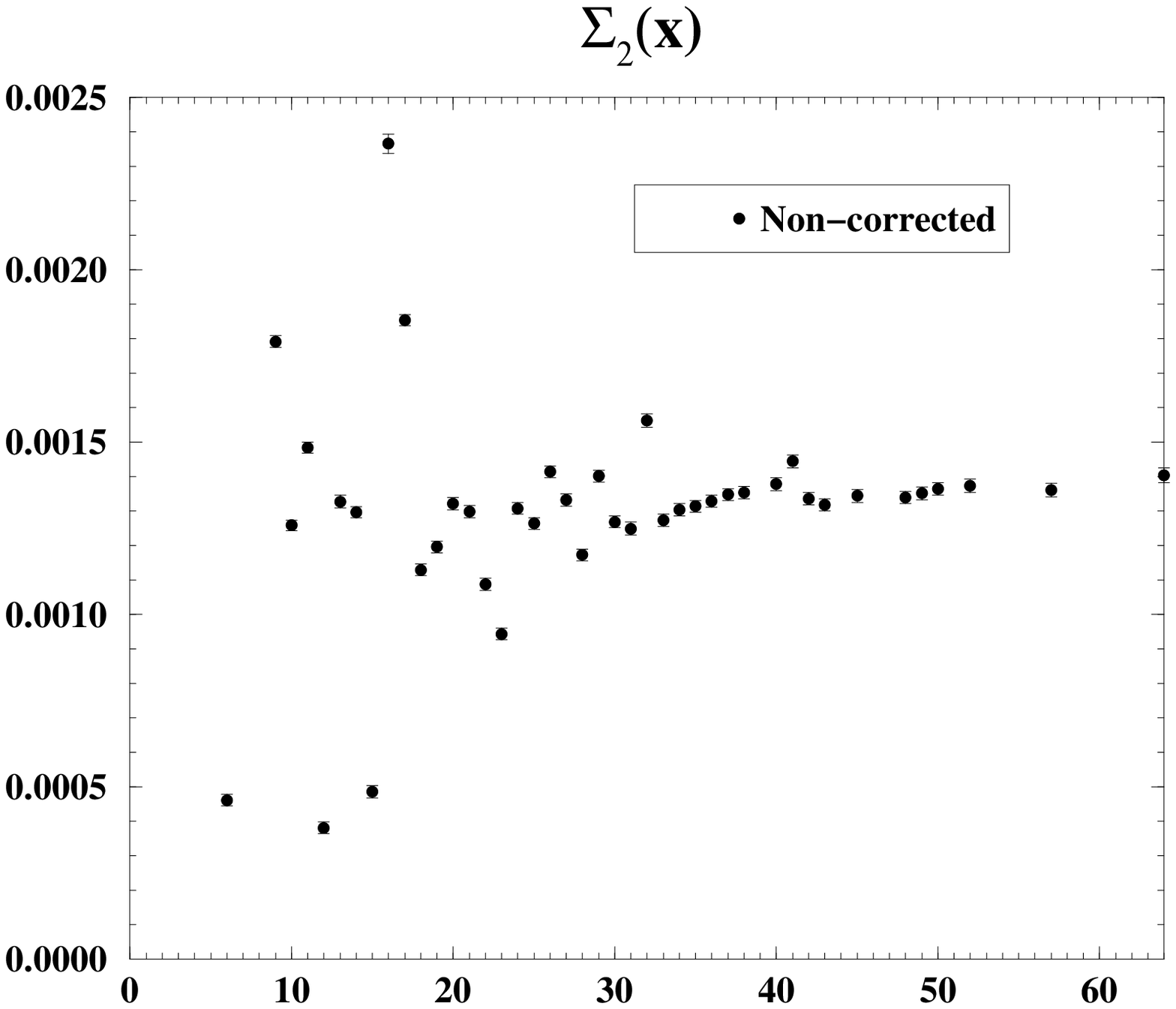}
\\ \vspace{-1.0cm} \epsfxsize5.4cm\epsffile{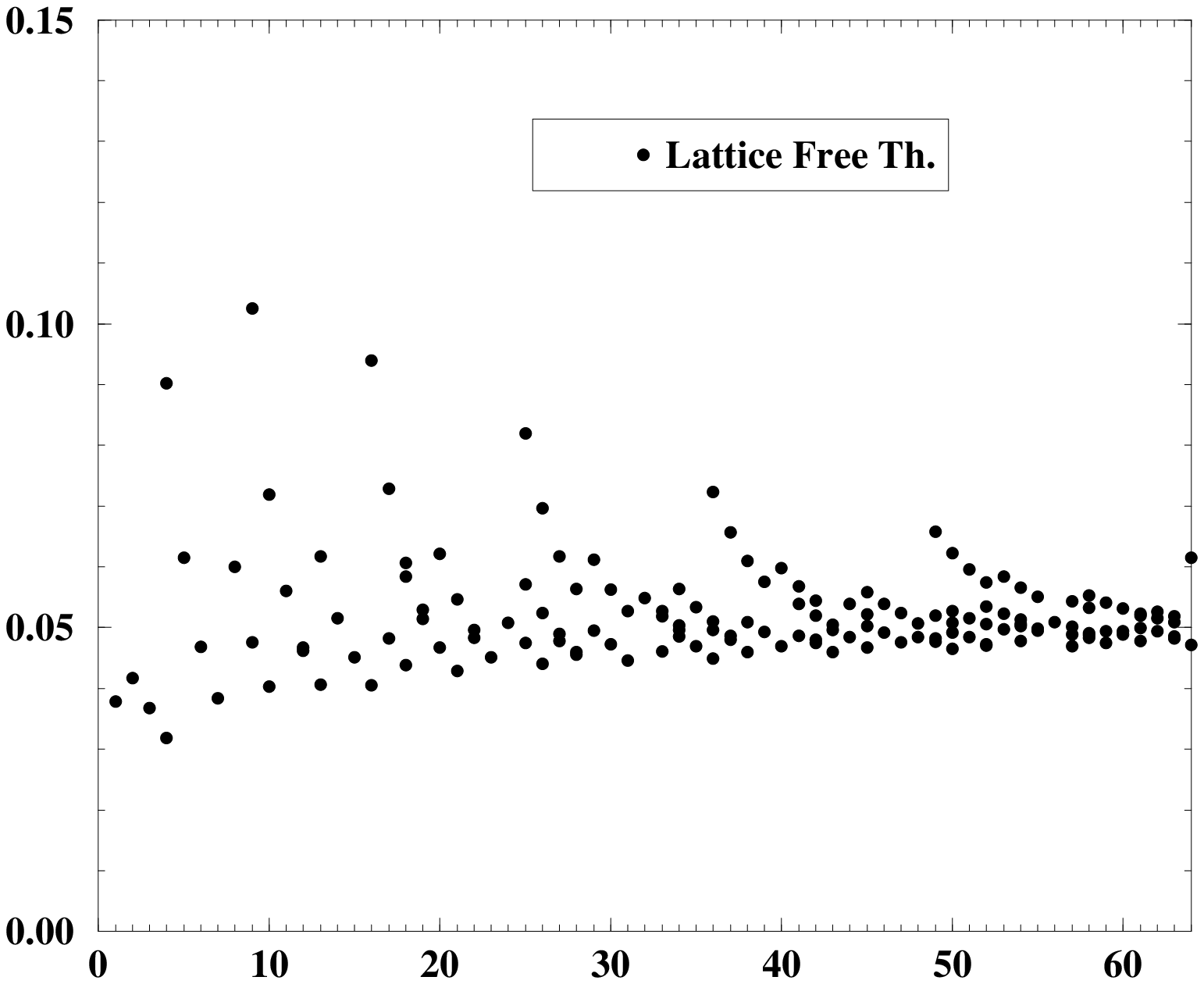} & \hspace*{0.5cm}
\epsfxsize5.7cm\epsffile{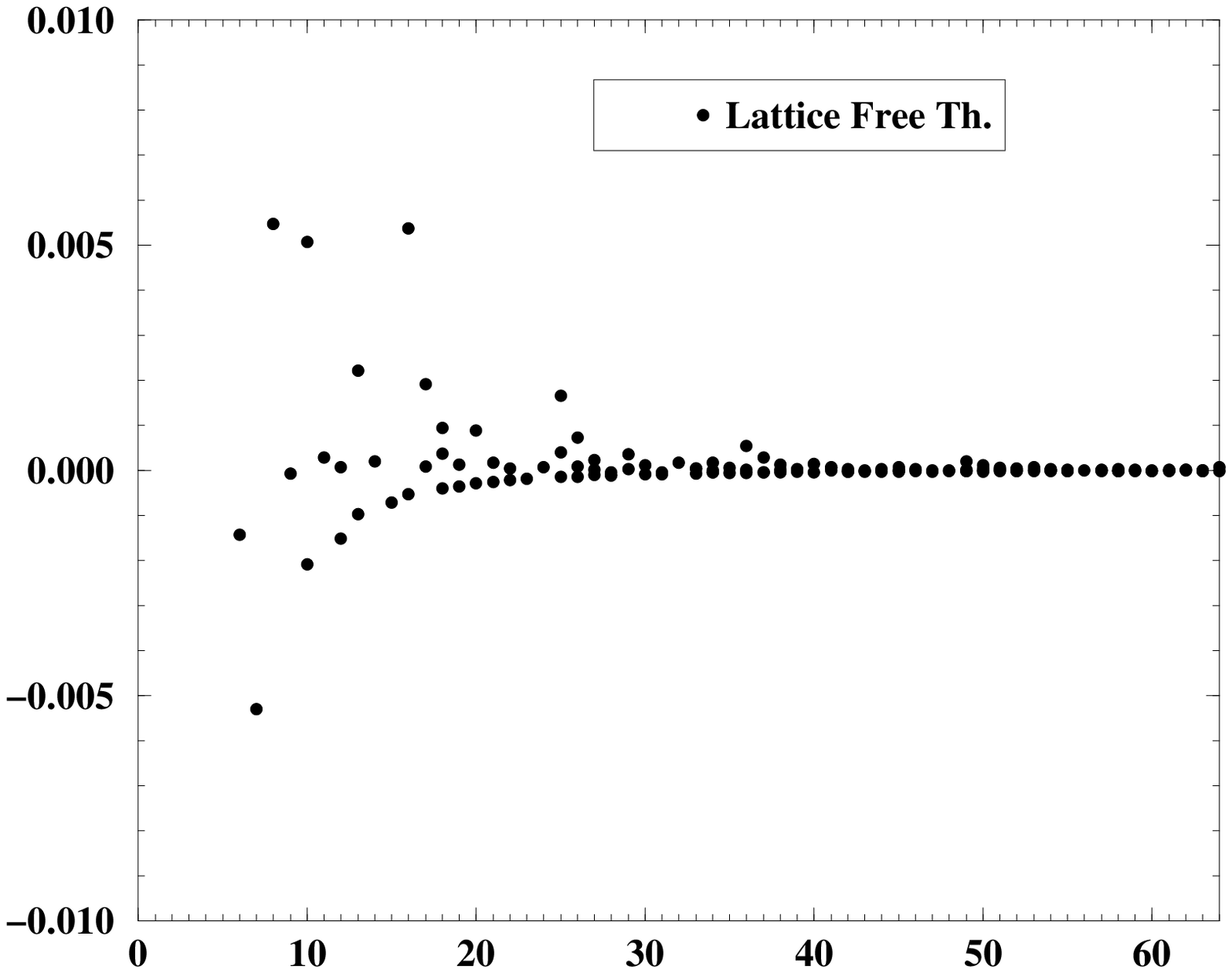} \\ \epsfxsize5.7cm\epsffile{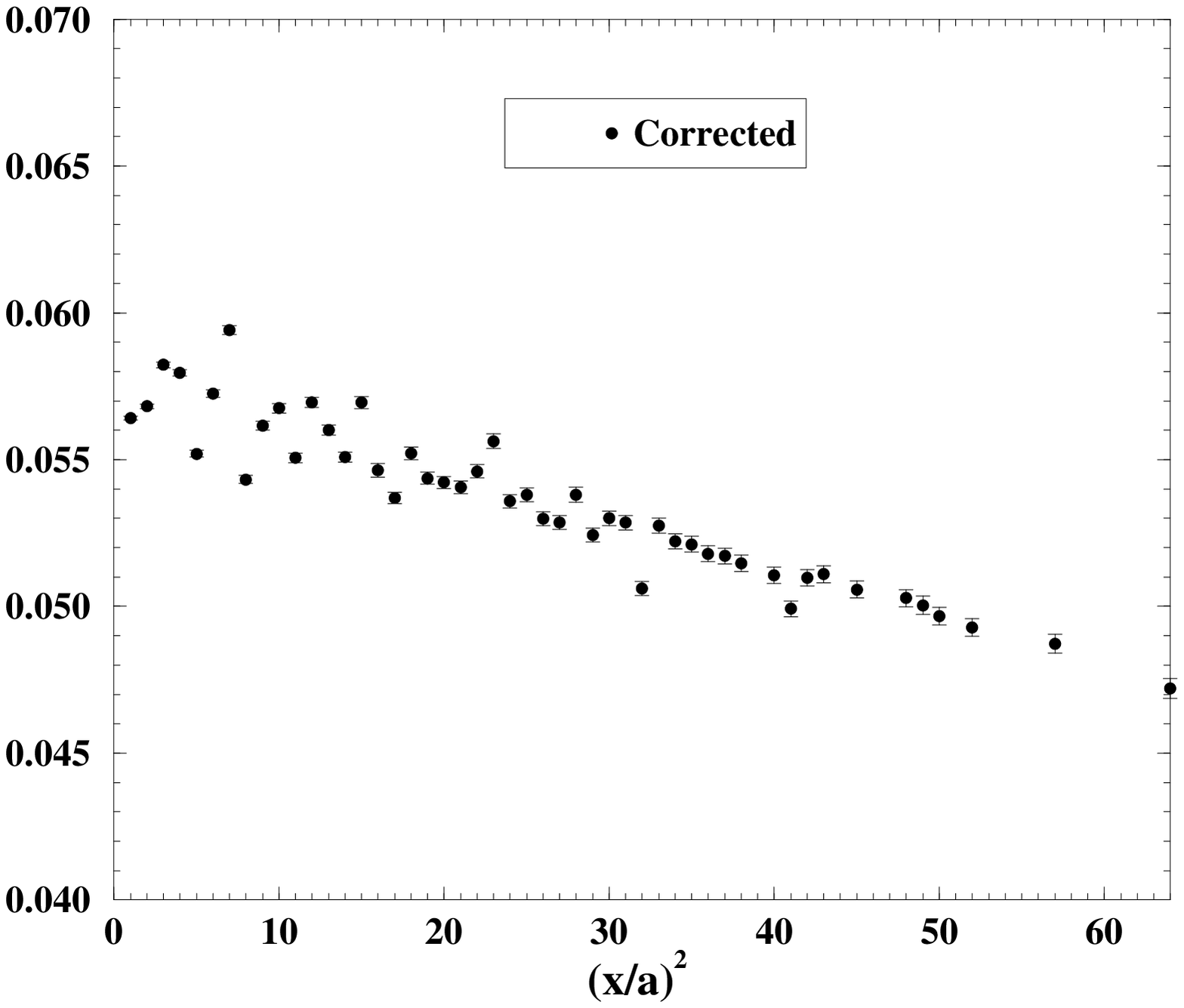} & \hspace*{0.5cm}
\epsfxsize5.7cm\epsffile{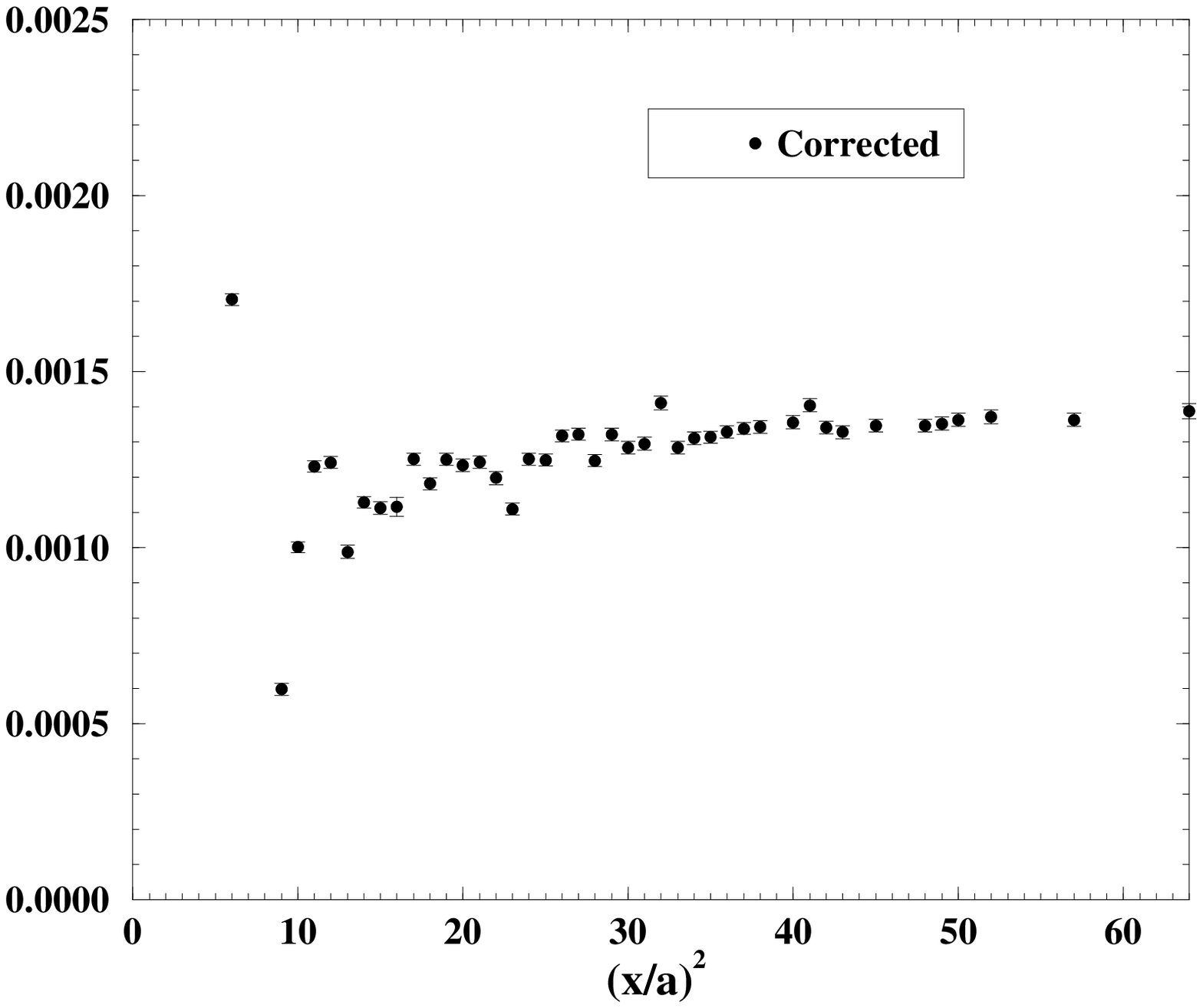}
\end{tabular}
\vspace{-0.5cm} \caption{\small\it The bare form factors $\Sigma_1$ (left panels) and $\Sigma_2$ (right
panels). From top to bottom: form factors in the interacting theory at k=0.1349; in the free lattice theory
at infinite volume and in the chiral limit; corrected form factors (as defined in the text). }\label{fig1}
\end{figure}
 The quantities directly evaluated from the numerical simulation are the scalar form factors
 of the quark propagator, defined from
 \be
S(x^2)= \frac{\slash x}{(x^2)^2}\, \Sigma_1(x^2)+\frac{1}{x^2}\, \Sigma_2(x^2). \label{eq:form} \ee
  We have calculated the Wilson coefficients $C_i(x^2)$ in eq.~(\ref{eq:ope})
   at ${\cal O}(\as)$ and resummed at the NLO in the $\ms$ scheme.
The NLO expressions of the coefficients and of the evolution functions can be found in ~\cite{Gimenez:0503001}. From
here on, $x$ indicates the euclidean distance in lattice units squared, $(x/a)^2$.

The scalar form factors $\Sigma_{1,2}(x,\mu)$ are calculated in the Landau gauge and shown in fig.~\ref{fig1}
(top) for $\kappa=$ 0.1349 as function of $x$. The points show fish-bone curves due to the spread of the
results obtained from different lattice sites which correspond to the same $x^2$ in the continuum limit,
especially at short distance. Similar patterns can be observed in the form factors computed in the free
lattice theory and plotted in fig.~\ref{fig1}(center). This suggests we are observing lattice artifacts. We
can reduce the discretization errors from ${\cal O}(a^2)$ to ${\cal O}(\as a^2)$ by correcting the lattice
data with the lattice artifacts evaluated in the free theory. In detail, we define \be \Sigma_1^{corr}(x)=
\left(\frac{\Sigma^{cont}_{1,\,free}(x)}{\Sigma^{lat}_{1,\,free}(x)}\right) \, \Sigma_1(x)
\,\qquad\qquad\Sigma^{cont}_{1,\,free}(x)=\frac{1}{2\pi^2}\, ,\label{eq:corr1} \ee \be
\hspace{-0.3cm}\Sigma_2^{corr}(x)=\Sigma_2(x)-\Sigma^{lat}_{2,\,free}(x)\hspace{1.6cm}\quad
\Sigma^{cont}_{2,\,free}(x)=0\, , \label{eq:corr2}\ee The result of the correction is
shown in fig.~\ref{fig1}(bottom). These are the data actually used in the extraction of the QC.

\section{Renormalization}
\begin{figure}[t]
\begin{center}
\includegraphics[width=7cm]{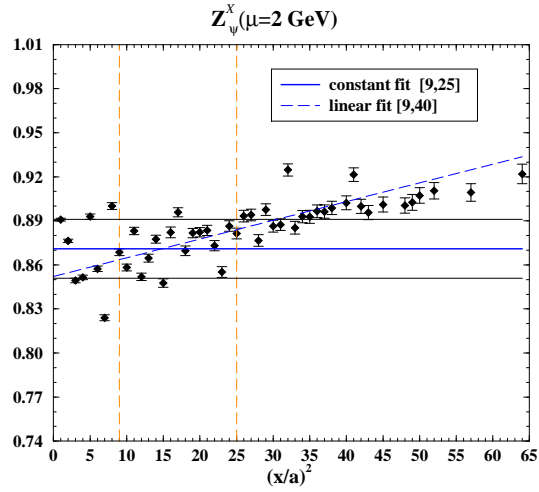}
\end{center}
\vspace{-0.5cm} \caption{\small\it Values of $Z_\psi^X(\mu=2 \gev)$ for different values of $x\equiv
(x/a)^2$. The solid lines indicate the results obtained from a constant fit in $x$ and the estimated
systematic error. The dashed vertical lines show the $x$-range where the constant fit is performed. The
result of a linear fit is also shown (dashed line); the estimate of $Z_\psi^X(\mu=2 \gev)$ is given by the
intercept.} \label{fig2}
\end{figure}

 As anticipated, we need to compute only the renormalization constant of the quark field. For this purpose, we
use the non-perturbative renormalization scheme named $X$-scheme in ~\cite{Gimenez:2004me}-~\cite{xspace} and
defined by the condition\be Z_\psi^X(\mu=1/x)\,\Sigma_1(x) = \Sigma^{cont}_{1,\,free}(x) \label{eq:Xsp}\ee in
the Landau gauge and in the chiral limit. From eqs.(\ref{eq:ope}) and (\ref{eq:form}),\be
\Sigma_1(x)=\frac{1}{2\pi^2}\,C_I(x)+\cdots \,.\ee In order to reach the chiral limit, linear and quadratic
chiral extrapolations have been considered and the differences are included in the systematics. The result
for $Z_\psi^X(\mu=1/x)$ as obtained from eq.~(\ref{eq:Xsp}) is shown in fig.~\ref{fig2}. The final estimate
of the renormalization constant is obtained from a constant fit in the range [9,25]: \be Z_\psi^{X}(2\gev) =
0.871\pm 0.003_{stat}\pm 0.020_{syst} \label{zpsi} \ee This is also the of value of $Z_\psi^{\ms}(2\gev)$, in the Landau
gauge. A fit which includes a long distance region, $x\in[9,40]$, has been also performed in order to
evaluate the systematics.

\section{Determination of the quark condensate}
From eqs.(\ref{eq:ope}) and (\ref{eq:form}) \be
\Sigma_2^\ms(x,\mu)=\frac{1}{4\pi^2}\,C_m^\ms(x,\mu)\,m^\ms(\mu)-\frac{1}{4N_c}\,C_{\bar\psi\psi}^\ms(x,\mu)
\,\langle\bar\psi\psi\rangle^\ms(\mu) \, x^2+\cdots  \ee
 Therefore in the chiral limit the QC represents the leading term in the OPE of $\Sigma_2(x,\mu)$.
 A linear and a quadratic extrapolations have been performed in order to evaluate the systematics. 

\begin{figure}[t]
\vspace{0.5cm}
\begin{center}
\includegraphics[width=7cm]{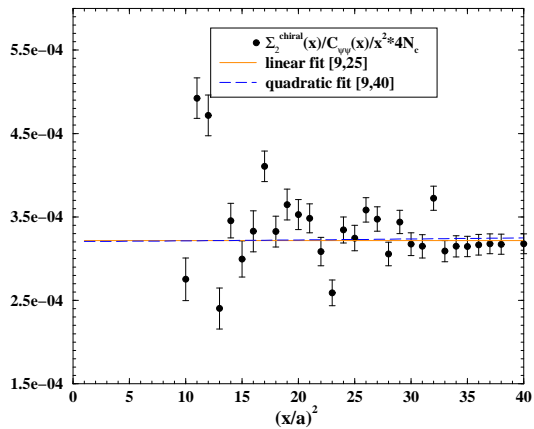}
\end{center}
\vspace{-0.5cm} \caption{\small\it Values of $\frac{(\Sigma_2^{\ms}(x,\mu))^{\rm chiral}}{C_{\bar\psi\psi}
(x,\mu)\,x^2/4 N_c} = -\langle
\bar\psi\psi \rangle^\ms(\mu) + {\cal O}(x^2)$ as a function of $x\equiv (x/a)^2$. A constant 
and a linear fit in $x$ are shown. The absolute value of the quark condensate is
given by the intercept.} \label{fig4}
\end{figure}

 In order to identify
directly the QC as the intercept, the fits in $x$ are performed on the normalized quantity \be 
 -\frac{(\Sigma_2^{\ms}(x,\mu))^{\rm chiral}}{C_{\bar\psi\psi} (x,\mu)\,x^2/4 N_c} = \langle
\bar\psi\psi \rangle^\ms(\mu) + {\cal O}(x^2) \label{eq:qI} \ee Some results are collected in
table~\ref{tab:cond} and the result of a constant fit in [9,25] is shown  in fig.~\ref{fig4}.

\begin{table}[t]
\begin{center}
\begin{tabular}{|c|c|}
\hline Constant fit [9,25] & \hspace{0.8cm}Linear fit [9,40]\hspace{0.8cm}  \\ \hline
$-(265\pm 5)^3$   & $-$ \\
$-(266\pm 4)^3$   & $-(265\pm 7)^3$ \\
\hline
\end{tabular}
\end{center}
\vspace{-0.5cm} \caption{\small\it Values of the chiral quark condensate in the $\ms$ scheme at the scale
$\mu=2 \gev$.} \label{tab:cond}
\end{table}
As a check of the result we also extract the QC from $\Sigma_2(x,\mu)$ by inverting the order of the fits:
first, the fit in $x$ including the mass term, then the chiral fit. We obtain a perfect agreement of the
results with the central values given in table~\ref{tab:cond}. We note, however, that in order to stabilize
the results of the fit, we have not been able to treat the quark masses as free parameters, but their values
have been kept fixed to the ones determined in ref.\cite{masse}.
Averaging the results and adding the systematic error we obtain
\begin{equation}
\langle\bar \psi \psi \rangle^{\ms}(2 \textrm{GeV})=\condval \,.
\end{equation}
The systematic uncertainty amounts
 to 26\%, the main contribution being the spread of the points in the fitting regions (18\%).  The reason is
 that the fitting intervals have been chosen in the region $x\in[9,40]$ in order to satisfy the condition
\be 1 < \sqrt{x} \simle 1/(a\Lam) \quad \quad\quad x\equiv\left(\frac{x}{a}\right)^2\label{eq:window} \ee
 where the discretization errors are under control (lower limit) and  the perturbative calculation of the
coefficients at a scale $\mu\sim 1/x$ is reliable (upper limit). But at the value of the lattice spacing
considered in the present simulation, this region is small and affected by non-negligible discretization
effects, even after the correction of the lattice artifacts described in sec.\ref{2} has been applied.

We estimate that neglecting higher order terms in the OPE introduce an uncertainty of $\sim$5\%.  All the
other sources of systematic error are due to the lattice uncertainties: 15\% the error due to the
determination of the lattice spacing, 10\% the quenched estimate of $\Lambda_{QCD}$,  9\% the use of
linear/quadratic chiral extrapolations, 5\% the choice of the fitting interval, $<$1\% the error on
renormalization constant of the quark field and the use of vector or axial definitions of the quark masses.
We could not quantify the finite volume effects, which should be small in the fitting regions (see
~\cite{Gimenez:2004me} for a study in the free theory), and, above all, the quenching approximation.


\end{document}